\documentstyle[prb,aps,epsf,floats]{revtex}
\newcommand{\beq}{\begin{equation}}
\newcommand{\eeq}{\end{equation}}
\newcommand{\beqr}{\begin{eqnarray}}
\newcommand{\eeqr}{\end{eqnarray}}
\newcommand{\e}{{\epsilon}}

\newcommand{\lol}{({L\over \lambda})}

\def\bp{{\mathbf p}}

\def\br{{\mathbf r}}

\def\br{{\mathbf r}}

\def\bS{{\mathbf S}}

\def\bz{{\mathbf z}}
\def\ba{{\mathbf a}}

\def\barg{{\bar{G}}}

\def\hO{{{\hat O}}}

\newcommand{\sigmab}{\mbox{\boldmath $\sigma $}}

\def\lam{\lambda}

\def\half{{1\over2}}

\def\2third{{2\over3}}

\def\eqa{\begin{eqnarray}}
\def\eea{\end{eqnarray}}

\def\a{{\alpha}}

\def\d{{\delta}}

\def\dbar{{\bar {\delta}}}

\def\pr{{Phys. Rev.}}
\def\prl{{Phys. Rev. Lett.}}
\def\prb{{Phys. Rev. {\bf B}}}

\begin{document}
\draft \flushbottom \twocolumn[
\hsize\textwidth\columnwidth\hsize\csname
@twocolumnfalse\endcsname
\title{Large spin-orbit effects in small interacting quantum dots}
\author{Ganpathy Murthy$^1$, R. Shankar$^2$}
\address{$^1$Department of Physics and Astronomy, University of Kentucky,
Lexington KY 40506-0055\\ $^2$ Department of Physics, Yale
University, New Haven CT 06520 }
\date{\today}
\maketitle
\begin{abstract}

We consider small ballistic quantum dots weakly coupled to the leads
in the chaotic regime and look for significant spin-orbit effects. We
find that these effects can become quite prominent in the vicinity of
degeneracies of many-body energies. We illustrate the idea by
considering a case where the intrinsic exchange term $-J{\mathbf S}^2$
brings singlet and triplet many-body states near each other, while an
externally tunable Zeeman term then closes the gap between the singlet
and the one of the triplet states (with spin projection parallel the
external field). Near this degeneracy, the spin-orbit coupling leads
to a striking temperature dependence of the conductance, with
observable effects of order unity at temperatures lower than the
strength of the spin-orbit coupling. Under favorable circumstances,
spelled out in the paper, these order unity effects in the conductance
persist to temperatures much higher than the spin-orbit coupling
strength.  Our conclusions are unaffected by the presence of
non-universal perturbations.  We suggest a class of experiments to
explore this regime.

\end{abstract}
\vskip 1cm ]

\section{Introduction}
\label{intro}

Electrons in planar quantum dots experience spin-orbit
interactions due to intrinsic (Dresselhaus\cite{dressel}) terms
and due to the electric field arising from the confinement to two
dimensions (the Rashba term\cite{rashba}). Their strength is
measured by the parameter $L/\lambda$, where $L$ is the dot size
and $\lam$ is a spin-orbit scattering length. (We do not worry in
this paper about the two different scattering lengths for the two
terms.) Spin-orbit couplings have been thoroughly investigated in
noninteracting quantum
dots\cite{bert-ady,af,cremers,BCH,koneman,falko,ahmadian}, and
more recently, in interacting dots as well in certain special
regimes\cite{gorokhov,alhassid2,murthy}. Spin-orbit terms violate
spin  conservation and appear to move the symmetry class from
orthogonal to symplectic to first order in $L/\lambda$.  Aleiner
and Fal'ko (AF)\cite{af} used a judicious unitary transformation
to show that despite appearances the spin-orbit interaction enters
the hamiltonian only to order $(L/\lambda)^2$ and beyond. They
also showed that to order $(L/\lambda)^2$ (but not beyond) a new
quantity $\sigma_{z}^{AF}$ (similar algebraically to the usual
$\sigma_z$ but distinct from it) is conserved.  The crossover to
the new symmetry class occurs at a scale $\e^{AF}\simeq(L/\lam)^4
g\dbar$, and the violation of $\sigma_{z}^{AF}$ occurs only at the
scale\cite{af} $(L/\lam)^6 g\dbar$, where $\dbar$ is the mean
single-particle level spacing, $g\dbar$ is the Thouless energy,
and $g$ is the Thouless number. Thus in a small dot with $L\simeq
200 \ nm$ and $\lambda \simeq 2000-3000 \  nm$, spin-orbit effects
are expected to be extremely small.

Here we ask how these considerations, derived for non-interacting
particles, are modified if the exchange interaction ($-J\bS^2$) of the
Universal hamiltonian\cite{univ-ham,hu-reviews} is included.  Recall
that the Universal Hamiltonian is the correct low-energy fixed
point\cite{mm} deep inside the Thouless band for sufficiently weak
interactions\cite{sm}. We find that in the presence of the exchange
coupling $J$ the spin-orbit effects really are of order ${L\over
\lambda\sqrt{g}}$ in the hamiltonian and that in the proximity of
many-body degeneracies (induced largely by $J$ and aided by a Zeeman
field which also serves as a knob for dialing through the transition),
effects {\em linear} in $L/\lambda$ may be found in energies, and most
strikingly, effects {\it of order unity} may generically be found in
the conductances (provided $T$ is low enough). This paper focuses on
calculating these effects, and proposing experiments to measure
them. The optimal situation is in dots where low temperatures (in
units of level spacing) may be attained without too much reduction of
${L\over \lambda}$. Currently, for dots of linear size $L\simeq
200nm$, the typical mean level spacing is $\bar{\delta}\simeq 70-100
\mu eV$ and temperatures of $T\simeq .1 \bar{\delta}$ are
achievable. The effects under study will be more readily detected if
$T/\bar{\delta}$ is lower by a factor of 5 or more.

In Section II we consider the Universal Hamiltonian plus a
spin-orbit term. We review some previous work and elucidate how
and when various powers of $L/\lambda$  appear in observables and
how the presence of a $J\ne 0$ affects this. We then turn to our
main result, which is to show that in certain conductance
measurements
 there is a good chance of seeing spin-orbit effects of order
unity. We estimate the effects in terms of a single parameter and
propose experiments to reveal them and to interpret the
data. Discussions and conclusions follow in Section III.

\section{Low temperature  physics}

Our starting point is the hamiltonian

 \beq H= H_U+ H_{so}+H_Z. \label{hamiltonian}
\eeq

The Universal Hamiltonian\cite{univ-ham} for the case of
orthogonal symmetry is 

\beq H_U= \sum_{\mu  s}\e_{\mu}\
c^{\dagger}_{\mu  s}c^{}_{\mu s}\ +{U\over 2}N^2
-J\bS^2.
\eeq 

Here $c^{\dagger}_{\mu  s}$ creates an electron
in a single-particle state $\mu$ of energy $\varepsilon_{\mu }$
and spin projection $s_z=s/2$ in a chaotic but ballistic
dot. The charging energy is $U$, $N$ is the particle number, and
$\bS$ the total spin. The essence of the spin-orbit hamiltonian is
captured in first-quantization by the expression \beq H_{so}=
{\mbox{{$\bz\cdot\sigmab \times \bp $}}\over 2m\lambda }
.\label{soint}\eeq where $\bz$ is a unit vector along the
$z$-axis. For simplicity of discussion we keep the
Rashba\cite{rashba} term only.  The Zeeman term due to a parallel
field (chosen to be along the $x$-axis) is $H_Z=-E_Z S_x$.

We will be dealing with the case when the spin-orbit coupling is very
weak, in the sense that the Aleiner-Fal'ko crossover scale $\lol^4
g\dbar\ll\dbar$, which is applicable to small dots. The opposite
regime for large dots leads to quite different
physics\cite{alhassid2,murthy}, which we will comment on in the
conclusion.

Before plunging into details we outline our strategy. Imagine that
there is no spin-orbit coupling, that we are at very low temperatures
$T\simeq 0$, and are sitting at the Coulomb Blockade\cite{CB}
conductance maximum obtained by tuning the gate voltage $V_g$ to make
an $N$-body ground state with $(S=0, S_x=0)\equiv (0,0 )$ degenerate
with an $N+1$- body ground state with $(S=\half,S_x=\half )
\equiv(\half, \half )$. Now imagine turning on the Zeeman term $E_Z$
to cause a ground state transition in the $N$-body sector so that
$(S=1,S_x=1)\equiv (1,1)$ beats $(0,0)$. The $-J\bS^2$ term lowers
$E_{Z}^{*}$, the point where all three states-$(0,0),(1,1),(\half
,\half )$- become degenerate, by bringing down all three triplet
states. At $E_{Z}^{*}$ the conductance will have a different value as
compared to either side where only two states are degenerate.  For
example (as will be seen in detail below) in the case where all
tunnelling widths are equal at both point contacts, the conductance at
$E_{Z}^{*}$ will be ${4 \over 3}$ of its value of on either side.

How does a spin-orbit interaction modify this expectation? By
eliminating the level crossing, and yielding a unique ground state
in the $N$-body sector, it will, in the illustrative example being
considered, kill the bump entirely. This is an example of the
order unity effect on the conductance due  the spin-orbit
interaction. If we now heat up the system so $T\simeq \alpha $
where $\alpha$    defines the scale of the spin-orbit interaction,
the bump in conductance, rounded by the elevated temperature, will
reemerge,  since the two  states that avoided each other will
contribute more or less equally. When we get down to details we
will find that while the above mentioned behavior can indeed occur
(provided all tunnelling rates  are equal  at both leads), so can
many others, some even more dramatic, {\em the actual behavior
being sensitively dependent on the wave functions at the leads}.
How then are we to predict what to expect? We shall see that with
a few additional conductance measurements one can obtain the
relevant information on  these wave functions and know in advance
what to expect of a given conductance peak in a given dot.

\subsection{Essential background}

Here we review some recent results on the  powers of the
spin-orbit coupling that should appear in noninteracting dots.

Let us begin with a simple non-interacting hamiltonian: \beq H=
{\bp^2\over 2m}+V(x,y)+ {\bz \cdot {\sigmab \times \bp}\over 2 m
\lambda} =H_0+H_{so}\label{AF0}\eeq where  $V$ is the confining
potential. The spin-orbit term, considered as a perturbation, is of
order $1/\lambda$ and we consider its impact on single-particle levels
described by time-reversal invariant real wave functions.  We will
refer to powers of ${1\over \lam} $ as powers of $L/\lambda$ since in
the end the $L$ in the numerator will invariably appear in physical
quantities. For future reference, we define the orthogonal basis of
$H_0$ 
\beq
H_0={\bp^2\over2m}+V(x,y)
\eeq

Going over to second-quantized notation we can express $H_0$ in its
own orthogonal eigenbasis in the many-body Fock space as
\beq
H_0=\sum\limits_{\mu,s} \e_\mu c^{\dagger}_{\mu,s}c_{\mu,s}
\eeq
and the total spin operator as $\bS=\half\sum
c^{\dagger}_{\mu,s}{\sigmab_{ss'}}c_{\mu,s'}$.

Because $H_{so}$ has no diagonal matrix elements in the basis of the
unperturbed hamiltonian, we expect this term to appear to order
$\lol^2$ in the energy shift of levels. It was pointed out by
AF that this is illusory\cite{af}. Upon completing squares they bring the
kinetic term to a form \beq {(\bp + {\bz \times \sigmab \over 2
\lambda})^2\over 2m}-{1\over 4m\lam^2}\label{AFH} \eeq
 when it
becomes apparent  that the particle is coupled to a spatially
constant nonabelian vector potential. Using the gauge
transformation \beq U= \exp \left[ i \bz \cdot \br \times \sigmab
\over 2 \lambda \right] \label{unit}\eeq and expanding  it in
inverse powers of $1\over \lam$, they transform to new variables
in terms of which the kinetic term becomes  \beq H_{AF}={(\bp_{AF} - {1\over
2}\sigma^{AF}_z \ba_{\perp})^2\over 2m}-{1\over 4m\lam^2}\eeq where
\beq \ba_{\perp} = {\br \times \bz \over 2 \lambda^2}.\eeq Terms
of order $\lol^3$ and higher are not shown.

Note that $\sigma^{AF}_z$ and $\bp_{AF}$ are not the same as
$\sigma_z$ and $\bp$, while $\br$ remains untransformed because it commutes with $U$. We can
rewrite $H_{AF}$ as an unperturbed orthogonal Hamiltonian $H_{0,AF}$ (the first two terms in the following equation)
perturbed by spin-orbit terms
\beq
H_{AF}={\bp_{AF}^2\over 2m} +V(\br)+H_{so,AF}
\eeq
Going over to second quantization, we express $H_{0,AF}$ in its orthogonal eigenbasis
\beq
H_{AF}=\sum\limits_{\mu,t} \e_{\mu,AF} d^{\dagger}_{\mu,t}d_{\mu,t} +H_{so,AF}
\eeq
where we have denoted the new fermionic operators $d,\ d^{\dagger}$ to
 emphasize their difference from the earlier $c,\ c^{\dagger}$'s.  We
 now define a total spin operator in the $H_{0,AF}$ eigenbasis as
 $\bS_{AF}=\half\sum
 d^{\dagger}_{\mu,t}{\sigmab_{tt'}}d_{\mu,t'}$. Clearly,
 $\bS_{AF}$ is not the same operator as $\bS$. In fact,
\beq
\bS=U^{\dagger}\bS_{AF}U=\bS_{AF}+\delta\bS_{AF}
\label{spin-transform}\eeq
which shows that $\bS_{AF}$ is some complicated combination of the
original spin and orbital degrees of freedom.

Note that the vector
 potential is not fully gauged away. This must be so since a spatially
 constant nonabelian vector potential can have a nonzero field
 strength, which in our case is $f_{\mu
\nu}=\left[ A_{\mu}\ , \ A_{\nu}\right]\simeq \sigma_z/\lambda^2$.
This is why in the end the  particle is
 coupled to a magnetic field of size $\lol^2$ and a sign
which depends on $s_z$. This coupling  will contribute to energies
or rates to order $\lol^4$. It is also clear that $\sigma_z$ in
the new AF basis (to be referred to as $\sigma_{z}^{AF}$) is
conserved by the $\ba_{\perp}$ term and violated only by higher
order terms.

It is instructive to ask how this correct physics would manifest
itself if one did not make the AF transformation and worked
perturbatively in the original orthogonal basis. The heart of the
equivalence between the two bases is the operator identity \beq
{\bp\over m}={i\over\hbar}[H_0,\br]\label{p=rdot}\eeq which means
that

\beq
\langle \mu |\bp|\nu\rangle={im\over\hbar}(\e_\mu-\e_\nu)\langle \mu |\br
|\nu\rangle
\label{p=rdot2}\eeq

This identity was emphasized  and exploited by Halperin {\it et
al}\cite{bert-ady} in their analysis of the interplay between
spin-orbit and Zeeman couplings.  To see explicitly how this
enters our discussion, consider energy level shifts of order
$\lol^2$ in the orthogonal basis. One finds for the second order
energy shift in the single-particle state $\mu $ of either spin:

\begin{eqnarray} \Delta E^{(2)}_\mu
&=& {1\over 4 m^2 \lam^2} \sum_{\nu\ne \mu} {\langle \mu | \bp |\nu\rangle\cdot
\langle \nu | \bp
|\mu\rangle \over \epsilon_\mu - \epsilon_\nu} \\
&=& {1\over 4 m^2 \lam^2}{im\over \hbar}  \sum_{\nu\ne \mu}{ \langle \mu
| \bp |\nu\rangle \cdot\langle \nu |
\left[H_0,\ \br\right] |\nu\rangle \over \epsilon_\mu - \epsilon_\nu}\\
&=& -{1\over 4 m^2 \lam^2}{im\over \hbar}\left(\langle \mu|\bp \cdot
\br |\mu \rangle +\langle \mu |\br |\mu\rangle \cdot  \langle \mu |\bp
|\mu\rangle
\rangle\right)\\
&=& -{1\over 4 m^2 \lam^2}{im\over \hbar}\langle \mu|\bp \cdot \br
|\mu \rangle \label{sum}
\end{eqnarray}
where  we
have used the time-reversal\footnote{If one denotes a state by
$|\psi\rangle$ and its time-reversed state by $|\psi^\tau\rangle$,
and if one denotes the time-reversed operator corresponding to
$\hO$ by $\hO^\tau$, then we have
$\langle\psi_1|\hO|\psi_2\rangle=\langle\psi_2^\tau|\hO^\tau|\psi_1^\tau\rangle$,
and
$\langle\psi_1|\hO_1\hO_2|\psi_2\rangle=\langle\psi_2^\tau|\hO_2^\tau\hO_1^\tau|\psi_1^\tau\rangle$.
Also note that $\br^\tau=\br$, and $\bp^\tau=-\bp$.} invariance of
the orthogonal states $|\mu\rangle$ to set $\langle \mu |\bp
|\mu\rangle=0$. Using time-reversal invariance again we get \beq
\langle \mu|\bp\cdot\br|\mu\rangle=-\langle
\mu|\br\cdot\bp|\mu\rangle=-\half\langle
\mu|[\bp,\cdot\br]|\mu\rangle \eeq ending up with a shift
$-{1\over 4m\lambda^2}$ {\em which is the state-independent shift
 obtained by completing squares.}

At fourth-order $\lol^4$ one will find corrections that are state
-dependent, in agreement with what is expected in the AF basis
based on $\ba_{\perp}$. Note that to get the answer correctly one
needs completeness. Had the second-order calculation been done in
a truncated space of low energy orbitals (say near the Fermi
energy) a nonexistent state-dependent shift would have been
obtained to order $\lol^2$. We mention this since we will perform
some calculations with such truncations and will need to be
careful.

All the above was for the case when the spin-orbit coupling was
constant  over the sample. However, as pointed out by Brouwer,
Cremers, and Halperin\cite{BCH}, since the Rashba term depends on
the confining electric field felt by the 2DEG, which in turn
depends on the density, one can fabricate additional gates over
the sample to make the spin-orbit coupling inhomogeneous. In this
case, the spin-orbit   matrix element has a typical size
$\lol\sqrt{g}$, and the system will crossover directly into the
fully symplectic ensemble at the energy scale $\lol^2 g\dbar$.

Let us now examine the effect of adding a $-J\bS^2$ term, which is
shown to be the correct starting point in the appendix. The result
is best seen in the AF basis. If we add a $-J\bS^2$ term to the
hamiltonian of Eq. (\ref{AF0}) in the original orthogonal basis, following Eq.
(\ref{spin-transform}) we end up with $-J(\bS_{AF}+\delta \bS_{AF})^2$ where
$\delta \bS_{AF}$ is the leading correction due to the action of
$U$. The leading term $ -J(\bS_{AF})^2$ commutes with
$\sigma_{z}^{AF}$ but the corrections do not. They will typically
produce effects of order $\lol^2{1\over g}$, (say in relaxation of
$S_{z}^{AF}$ ) but the effects can be linear in
$\lol{1\over\sqrt{g}}$ at and near degeneracies in many-body
states as will become clear soon.

The effect of $J$ may also be seen in the orthogonal basis, but not so
easily. If we turn on a $J$ and resort to many-body perturbation
theory to order $\lol^2$, $J$ will enter energy denominators, in a
calculation similar to the one leading to Eqn.  (\ref{sum}). If we set
$J=0$ in any of these results, we must recover the free-particle
case. Expanding the denominators in powers of $J/(\epsilon_\mu
-\epsilon_\nu)$ will generate nontrivial effects proportional to powers
of $J$. These will match the results described above in the AF basis.

We now turn to our main purpose.

\subsection{Very low temperatures}

We will begin by defining the states of interest in the extremely
low energy sector, $\Gamma << T< \alpha << \delta$, where $\Gamma$
is a typical single-particle level width due to coupling to the
leads.  This will allow us to get a firm grasp of all the relevant
effects. Later we will incorporate the effects of $T\simeq \a
<<\delta$. Let us first turn off $H_{so}$ and $E_Z$ in Eqn.
(\ref{hamiltonian}). Consider the $N$ particle state. Figure 1
shows the lowest energy state with $(S,S_x)=(0,0)$. It has two particles
in an orbital $\phi_0$ at the Fermi energy (set to zero). The
filled sea underneath can be ignored. The energy of this state is
\beq E_{0,0}= 2*0 +{U\over 2}N^2 - 0*J.\eeq We could also form an
$S=1$ state, symmetric in spin and antisymmetric between the
orbital $\phi_0$ and the one just above it with a spacing
$\delta$, labelled $\phi_1$.  The energy of these states is \beq
E_{S=1,S_x=\pm 1,0}= \delta +{U\over 2}N^2 - 2J.\eeq Note that
$\delta - 2J$, the energy difference between singlet and triplet,
can be much smaller than $\delta$. This is why we keep the orbital
$\phi_1$ but not higher ones. For now, when we do not have the
Zeeman term, we will ignore the triplet state, assuming $\delta -
2J>>T$.

\begin{figure}
\narrowtext \epsfxsize=2.4in\epsfysize=2.0in \hskip
0.3in\epsfbox{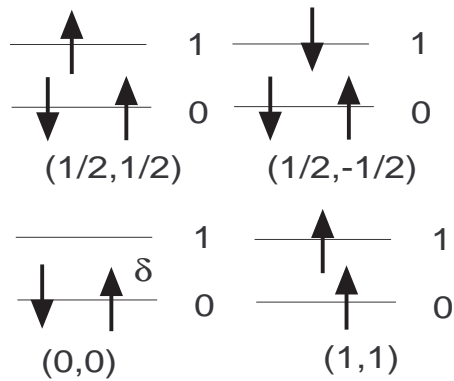} \vskip 0.15in \caption{Schematic of
obitals and four of the low energy states (labelled by $(S,S_x)$)
that play a key role at $\Gamma<<T<< \a<<\delta$. The arrows
denote spins.} \label{orbitals}
\end{figure}

Now for the $N+1$ particle state. The  extra particle will go to
the orbital state $\phi_1$ with spin up or down, the two states
being degenerate. The energy of these states is \beq E_{\half, \pm
\half}= \delta +{U\over 2}(N+1)^2 -{3 \over 4}J.\eeq

By varying a gate voltage $V_g$ we can make the $N$ and $N+1$
particle states degenerate so the system conducts. Assume this has
been done. Let us calculate the conductance of this
 peak, using simple ideas. Later we turn to a master formula, based on
rate equations\cite{beenakker,alhassid1}, that covers any situation
where $T>>\Gamma$.

   We want an electron to jump in from the left with $s_x$-independent
width $\Gamma_{0,\half}^{L}$ and hop out of the right with
$s_x$-independent width $\Gamma_{0,\half}^{R}$. The subscript on
$\Gamma$ labels the $S$ values of the $N$ and $N+1$ particle
states degenerate at this peak. The dot has to be in the singlet
N-particle state (and not in either of the doublet N+1- particle
states of the same energy) to receive this electron. Since all
three states are degenerate at this peak, the probability for this
is $1/3$. The reduced conductance $\bar{G}$ (which has a trivial
temperature dependence removed)  is given by \beq  \bar{G}_{0\to
\half}^{0} \equiv {2 \hbar k T \over e^2}G= {2 \over
3}{\Gamma_{0,\half}^{L}\cdot \Gamma_{0,\half}^{R}\over
\Gamma_{0,\half}^{L}+ \Gamma_{0,\half}^{R}}.\label{reducedG}\eeq
(The logic behind this definition of $\barg$ will be apparent
later when we do a full-blown calculation of conductance.)  The
subscript on $\bar{G}$ stands for the two states that are
degenerate at this peak and the superscript reminds us that the
Zeeman term is absent: $E_Z=0$. The factor of $2$ comes from sum
over spins. The dependence on $\Gamma$'s maybe understood as
follows. The left and right contacts are like {\em resistors} in
series while the $\Gamma $'s represent their {\em conductances}.
Thus we must add their reciprocals to obtain the reciprocal of the
effective conductance.

Since we want the electron to hop on the upper orbital described
by $\phi_1$,  \beq \Gamma_{0,\half}^{L/R}= \Gamma_{L/R}\
|\phi_1(L/R)|^2\label{gammaphi}\eeq where the arguments $L/R$
refer to the coordinates of the two leads within the dot and
$\Gamma_{L/R}$ represents an overall pre-factor for each contact.
For now let us  lump the overall scale $\Gamma_{L/R}$ into
$\phi_1(L/R)^2$ and write \beq \bar{G}_{0\to \half}^{0}\equiv
\barg^0= {2 \over 3}{|\phi_{1}(L)|^2\cdot |\phi_{1}(R)|^2\over
|\phi_{1}(L)|^2+ |\phi_{1}(R)|^2}.\eeq Later we will  return to
Eqn. (\ref{gammaphi}) that explicitly displays  the tunnelling
rate as a product of two factors, one  that depends on the point
contact and another  that depends on the dot wave function.

Let us now add a Zeeman term $E_Z$ and suitably adjust $V_g$ to
follow this peak (which has been done experimentally\cite{potok}).
Once $E_Z>>T$, only one of the spin states, with $S_x=+\half$ is
allowed in the dot. The reduced conductance now becomes \beq
\bar{G}_{0\to \half}^{1}\equiv \barg^1= {1 \over
2}{|\phi_{1}(L)|^2\cdot |\phi_{1}(R)|^2\over |\phi_{1}(L)|^2+
|\phi_{1}(R)|^2}\label{13}\eeq where the superscript on $\bar{G}$
signifies that the Zeeman term has a value strong enough to
essentially filter just one of the spin-$\half$ states, but not so
large the spin-1 state $(1,1)$ is competitive with $(0,0)$.
Compared to $G^0$, the pre-factor $\2third$ has become $\half$
since we have just one spin channel (not two as earlier) and
 two degenerate many-body states (not three as earlier).

 {\em The reduced conductance $\bar{G}$ has dropped
to a value equal to $3/4$ of the initial maximum:} \beq \barg^{1}
= {3\over 4} \barg^{0}.\eeq

\begin{figure}
\narrowtext \epsfxsize=2.4in\epsfysize=2.0in \hskip
0.3in\epsfbox{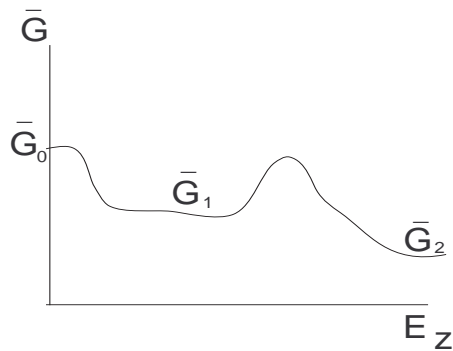} \vskip 0.15in \caption{Schematic of
$\barg$ as function of $E_Z$ at zero spin-orbit coupling ($\a=0$)
for very low $T$. Some rounding due to $T>0$ is assumed. The
initial drop from $\barg^0$ is due to suppression of one spin
orientation. The rise near $E_{Z}^{*}$ is due to the triple
degeneracy. } \label{smoothg}
\end{figure}

The situation is sketched   in  Figure (\ref{smoothg}).

Let us crank up the Zeeman term further till  the spin-1 state
$(1,1)$ meets the $(\half , \half )$ state and we  again have
three degenerate states. At this point denoted by a star, the
conductance is given by \beq \bar{G}_{}^* = {2 \over 3}
(\barg^1+\barg^2) \eeq where $\barg_2$ is the conductance past
this point where the spin-1 state $(1,1)$ is the ground state:
\beq \barg^2\equiv \bar{G}_{1\to \half}^{2}= {1 \over
2}{|\phi_{0}(L)|^2\cdot |\phi_{0}(R)|^2\over |\phi_{0}(L)|^2+
|\phi_{0}(R)|^2}\label{16} .\eeq Note that since the extra fermion
is placed in the lower orbital, only $\phi_0$ enters $\barg_2$.

To get the above result we need only realize that at the triple
degeneracy the two N-particle states each occur with probability
$1/3$ and the conductance via each is proportional to $\barg_1$ or
$\barg_2$.

The behavior near $E_{Z}^{*}$  depends on $\barg^1$ and $\barg^2$.
For instance if $\barg^1=\barg^2=\barg$, \beq \bar{G}_{}^* = {4
\over 3} \barg.\eeq In this case $\bar{G}$ first drops from
$\bar{G}^0$ to ${3\over 4}$ of its value to $\bar{G}^1$, rises to
$\bar{G}^0$ and drops to $\bar{G}^2$. If however $\bar{G}^1$ is
less than $\bar{G}^2/2$, the rise will be monotonic with no bump.
Likewise if $\bar{G}^2$ is less than $\bar{G}^1/2$, the fall will
be monotonic. Figure (\ref{smoothg}) describes the case
$\barg^1=\barg^2$.

For now let us focus on this case $\bar{G}^1=\bar{G}^2$, to get a
feeling for our approach, which is to ask how  the above mentioned
behavior of $\barg (E_Z)$ is altered by a spin-orbit term.  The
general form  of this term is  

\beq 
H_{so} = \sum_{\mu \nu} A_{\mu
\nu }\left[ c^{\dag}_{\mu \uparrow}c_{\nu \downarrow} +
c^{\dag}_{\nu \downarrow}c_{\mu \uparrow} \right] +iB_{\mu \nu}
\left[ c^{\dag}_{\mu \uparrow}c_{\nu \uparrow}- c^{\dag}_{\mu
\downarrow}c_{\nu \downarrow} \right] \label{M} 
\eeq 

where $\mu $
and $\nu $  labels single-particle orbital states, $\uparrow$ and
$\downarrow$ the direction of spin along $x$ and $A_{\mu \nu }$ or
$B_{\mu \nu}$ are real antisymmetric matrices: \beq A^T= -A \ \ \
\ \ \ B^T=-B.\eeq

The typical size of $A_{\mu \nu }$ or $B_{\mu \nu}$ can be
estimated to be $\lol \dbar/\sqrt{g}$ for the case of homogeneous
spin-orbit coupling, and $\lol \dbar\sqrt{g}$ for the case of
inhomogeneous spin-orbit coupling. For a typical small quantum dot
of size $200nm$ with $g\approx 10$, with the spin-orbit scattering
length being $2\mu m$, we find the spin-orbit matrix element to be
typically $0.03\dbar$ for the case of homogeneous spin-orbit
coupling, and $0.3\dbar$ for inhomogeneous spin-orbit
coupling\cite{BCH}. These estimates should be considered crude,
since the expressions for various matrix elements have prefactors
of order unity\cite{cremers} which depend on the geometry.

Let us now focus on the very low energy sector spanned by $(0,0),
(1,1) \mbox{ and}  (\half,\half )$. Since $H_{so}$ conserves
particle number, it does not mix the $S=\half$ states with $S=0,1$
and neither does it mix the two $S=\half$ states with each
other\footnote{In the case $S={3\over2}$ there would be an effect.
Also, in metallic grains the situation is more
complicated\cite{gorokhov}}. In the $S=0,1$ sector spanned by
$|0,0\rangle$ and $|1,1\rangle$ the effective hamiltonian matrix
assumes the form
\beq h_{so}=\left( \begin{array}{cc} 0 & \a \\ \a & 0 \\
\end{array}
\right) 
\eeq 
where the single parameter $\a\equiv A_{01}$ (see Eq.(\ref{M})) characterizes the
spin-orbit interaction. 

Note also that spin-rotation-invariant interactions beyond the
Universal Hamiltonian, such as Landau interactions,  will appear
in our basis  as diagonal corrections to the already random
energies.

This operator clearly leads to an avoided crossing and selects a
unique ground state in this sector. At the putative degeneracy
point the ground state with $\a\ne 0$, denoted by $|\a,*\rangle$
is unique and given by the antisymmetric combination: \beq
|\a,*\rangle = {\sqrt{\half}}\left[ |0,0\rangle
-|1,1\rangle\right] \eeq The reduced conductance at very low
temperature is \beq \barg^*(a)= {1\over 4} {\left[ |\phi_{1}(L)|^2
+|\phi_{0}(L)|^2\right]\left[ |\phi_{1}(R)|^2
+|\phi_{0}(R)|^2\right]\over \left[ |\phi_{1}(L)|^2
+|\phi_{0}(L)|^2+|\phi_{1}(R)|^2 +|\phi_{0}(R)|^2\right]}
\label{gofa}\eeq To derive this result we need to remember that
there is now a probability of $\half$ for the dot to be in the
N-particle state and that the widths are given, for example by
\beq \Gamma_{\half \to a}= \sum_{\sigma =\pm} \half |\langle
\mbox{{\small $\half,\ \half $}} |\psi^{\dag}_{\sigma
}(L)|0,0\rangle -\langle \mbox{{\small $\half,\ \half $}}
|\psi^{\dag}_{\sigma }(L) |1,1\rangle|^2\eeq and that in each of
the two terms only one value of $\sigma $ contributes.

If we are looking for the effects of $\a$, we need to compare
$\barg$ with and without $\a$. The answer depends on the wave
functions via $|\phi_{0/1}(L/R) |^2$.

Again consider the simple case where all are wavefunctions are equal
at both leads. It is readily seen that \beq \barg^*(\a)= \barg^1
=\barg^2,\eeq i.e., there is no second bump at the would-be level
crossing. This is an example of an order unity change in the
conductance due to a small spin-orbit perturbation. However, one
has to be at a temperature lower than $\alpha$ to see it.

Since the first bump is there at $E_Z=0$, the absence of the
second should be an unambiguous signal of $\a$. Furthermore as $T$
is raised beyond $\a$ (something we will consider in detail later)
the bump should reappear (though rounded) since the symmetric
combination of $|0,0\rangle$ and $|1,1\rangle$ (that got repelled
upwards) also starts contributing to $\barg$. The requisite $T$
will also tell us how large $\a$ is. While this is happening the
bump at $E_Z=0$ should be disappearing due to thermal effects.

This happy line of thought ends when we realize that in general
the four quantities $|\phi_{0/1}(L/R) |^2$ at the leads are not
all equal and that the ratio $\barg^*(\a)/\barg^*$ is very
sensitive to their values. We emphasize that even if we find
experimentally that $\barg^1=\barg^2$ we cannot say anything
definite.
 For example $\barg^1=\barg^2$  can also be realized if
 \begin{eqnarray}
 |\phi_{1}(R)|^2&=&|\phi_{0}(L)|^2 =A\\
 |\phi_{0}(R)|^2&=&|\phi_{1}(L)|^2=B.
 \end{eqnarray}
 In this case  we find
 \beq
 {\barg_{\half \to \a}\over \barg^1}= {\bar{G}_{\half \to \a}\over \barg^2}={1\over 4}
 (\sqrt{A/B}+\sqrt{B/A})^2\ge 1.
\label{symmcase} \eeq Thus the ratio can be any number $\ge 1$ with large ratios
resulting if either $A$ or $B$ is small. This behavior of
$\barg_{\half \to a}$ can be understood by examining Eqn.
(\ref{gofa}). It is seen that if tunnelling amplitude is weak at one
end and strong at the other for one wave function and the opposite is
true for the other, $\barg_{\half \to \a}$ is able to feed off the
bigger amplitude at both ends, unlike $\barg^1$ or $\barg^2$ which
necessarily have a weak amplitude at one end.

Thus if we are to say with confidence what effect $\a$ will have
on the ratio $\barg^*(\a )/\barg^*$ at $E_{z}^*$, we need to
extract the four numbers $|\phi_{0/1}(L/R) |^2$ in a given dot on
top of a given peak.  Here is our proposal.

 Recall from Eqn. (\ref{gammaphi} )that
 \begin{eqnarray}
 \Gamma_{0,\half}^{L/R}&=& \Gamma_{L/R}
|\phi_1(L/R)|^2\\
\Gamma_{1,\half}^{L/R}&=& \Gamma_{L/R} |\phi_0(L/R)|^2,
\end{eqnarray} where $\Gamma_{L/R}$ are {\em overall} prefactors
that control tunnelling rates at the left and right contacts and
are determined  by geometry, barrier heights and so on, and
independent of the wave functions. Let us take the given dot and
manually suppress tunnelling at the right contact by raising the
barrier there, so that $\Gamma^R<<\Gamma^L$. The conductances
assume the values $\barg^{1,2}(R)$ given by

 \begin{eqnarray}
 \barg^1 (R)&=& {\half} \Gamma_R|\phi_{1}(R)|^2\\
 \barg^2 (R)&=& {\half} \Gamma_R|\phi_{0}(R)|^2.
 \end{eqnarray}
 Their ratio gives us
 \beq
 r= {\barg^1(R)\over \barg^2(R)}= {|\phi_{1}(R)|^2\over
 |\phi_{0}(R)|^2}.
\label{rdefn} \eeq
 Likewise we can set $\Gamma^L<<\Gamma^R$ and measure
 \beq
 l={\barg^1(L)\over \barg^2(L)}= {|\phi_{1}(L)|^2\over
 |\phi_{0}(L)|^2}.
 \label{ldefn}\eeq

 Armed with these four numbers: $(\barg^1, \barg^2, r,l) $ {\em all
 measured at very low temperatures}, we can solve for $|\phi_{0/1}(L/R) |^2
 $:
 \begin{eqnarray}
|\phi_{1}(L) |^2&=& {2 \barg^1\barg^2(l-r)\over \barg^1-r\barg^2}\label{31}\\
|\phi_{1}(R) |^2&=& {2 \barg^1\barg^2(r-l)\over
\barg^1-l\barg^2}\label{32}\\
|\phi_{0}(L) |^2&=& {|\phi_{1}(L) |^2\over l}\label{phiol}\\
|\phi_{0}(R) |^2&=& {|\phi_{1}(R) |^2\over r}\label{phi0r}
\end{eqnarray}
While the right hand sides of the first two equations are not
positive definite (and can be negative if $\barg^1$ and $\barg^2$
are assigned arbitrary values) it can be shown  that if we input
any values for $\barg$ that come from a real measurement, i.e.,
given by Eqs. (\ref{13},\ref{16}) this will not happen. We will
illustrate this point later. Note also our assumption that the
wave functions in question are unaffected by what we do to alter
the tunnelling rates.

We can now express the conductances  of interest in terms of
knowns:
\begin{eqnarray}
\barg^*&=& {2 \over 3} (\barg^1+\barg^2)\\
\barg^*(\a)&=&  {\barg^1\barg^2 (1+r)(1+l)\over 2(\barg^1
+rl\barg^2)} \end{eqnarray}

It is now easy to specialize to any case we want. If for example
$\barg^1=\barg^2 =\barg$ and $l=r=1$,
 \beq
 \barg^*={4 \over 3} \barg^*(\a).
 \eeq

Before concluding this analysis of the $T<<\a$ case, we must ask
what the $({3\over
 2},{3 \over 2})$ level is doing in the meantime. In particular we
 do not want it to cross the $(\half ,\half )$ state before $(0,0)$
 crosses $(1,1)$, or better still before $\barg $ settles down to $\barg^2$.  To avert this we
 need at least
 \beq
  \delta'>\delta +J\eeq where $\delta'$ is the energy gap between the
orbital $\phi_1$ and the one just above it, $\phi_2$. Since for dots
with $r_s\simeq 1$ $J/\bar{\delta}\simeq .3$\cite{hu-reviews},
 we would be  fine in a system  where say $\delta \simeq .7
 \bar{\delta}$ and $\delta'\simeq 1.3 \dbar$. If however the gap
 at the Fermi energy is larger than the one above it, we should
 expect the $(\half, \half)\to ({3\over 2}, {3 \over 2})$
 transition to precede the $(0,0)\to (1,1)$ transition. The
 corresponding calculations are simply variants of the above.

 To test the picture described above we need to know where to
 look for putative degeneracies. Consider the peak associated with
 the $(0,0)-(\half,\half)$ degeneracy. The gate voltage $V_g$ (with leverage  suppressed) at which
 this occurs is
 \begin{eqnarray}
 V^{(0,0)-(\half ,\half)}_g(E_Z)&=& E_{\half , \half }-E_{0,0}\nonumber\\
  &=& {U\over 2}
 (2N+1)- {3\over 4}J - {E_Z\over 2}.\end{eqnarray}
 As a function of $E_Z$ it has a slope of $-\half$.
 On the other hand when $(1,1)$ wins we have
 \begin{eqnarray}
V^{(1,1)-(\half , \half )}_g(E_Z)&=& E_{\half ,\half
}-E_{1,1}\nonumber
\\
&=& {U\over 2}
 (2N+1)- \delta +{5\over 4}J + {E_Z\over 2}.\end{eqnarray}
which has  a slope of $+\half$ as a function of $E_Z$.

 At the point $E_{Z}^{*}= \delta -2J$, the slope changes from
 $-\half$ to $\half$, as shown in Figure ({\ref{vg}) \cite{falk} .

 If we turn on $\a$, we need to find $\epsilon_0(\a )$, the ground state of the energy
 matrix in the $(0,0)-(1,1)$ sector:
 \beq
 \left(%
\begin{array}{cc}
  0 &  \ \ \ \ \a \\
  \a & \ \  \mbox{{\small $\delta -\! 2J\!\!-E_Z$}}  \\
\end{array}%
\right) \eeq in terms of which  the gate voltage to stay on the
peak is \beq V^{\a-(\half,\half )}_g={U\over 2}(2N+1) - {3J\over
4} -{E_Z\over 2} -\epsilon_0(\a). \eeq Due to the avoided
crossing, $\epsilon_0(\a)$ is lower by an amount $\a$ and $V_g$ is
higher by an amount $\a$ at $E_{Z}^{*}$. A sketch is given in
Figure (\ref{vg}). Note that the results are most reliable near
$E_{Z}^{*}$ where things depend on $\a$. As we move away, results
depend on $\a^2$ and errors due to truncation of the orbital space
will enter. These are expected to be small for $\d-2J\ll\dbar$,
but we will not reproduce the arguments here.

\begin{figure}
\narrowtext \epsfxsize=2.4in\epsfysize=2.0in \hskip
0.3in\epsfbox{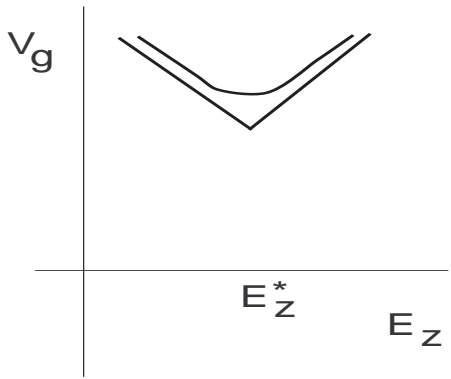} \vskip 0.15in \caption{Schematic of $V_g$ as
function of $E_Z$ at $T=0$. The straight and curved graphs
correspond to $\a=0$ and $\a\ne 0$ and approach each other
asymptotically. The slope change at $E_{Z}^{*}$ represents the
ground state transition $(0,0)\to (1,1)$.} \label{vg}
\end{figure}

 \subsection{Higher Temperatures:\ $\Gamma << T \simeq \delta
 -2J\ll\delta$} Now we consider a more detailed study of temperatures
 comparable to $\a$ but still considerably smaller than $\delta$. Thus
 we want to consider just six states: two at $S=\half$ and four at
 $S=0,1$.  In this larger space another parameter $\beta\equiv
 B_{01}$, associated with the matrix $B_{\mu \nu}$ of Eq. (\ref{M})
 enters.  This matrix element mixes the states $(S,S_x)=(0,0)$ and
 $(S,S_x)=(1,0)$. Since $(1,0)$ is split by about $\delta-2J$ near the
 triple crossing, we can estimate its effects to be of order
 $\beta^2/(\delta-2J)$ which is very small in the case of interest to
 us. We will therefore initially ignore it in the analysis.  Later we
 show a graph representative of its insignificance.

 As mentioned earlier, the cost of populating $\phi_1$  with $S=1$
 is not $\delta $ but $\delta - 2J$, and we assume $T$ is such that
many-body states  involving $\phi_0$ and $\phi_1$  are accessible
but not those
 involving higher orbitals.

The proper formalism for a finite $T$ computation involves rate
equations, developed by Beenakker\cite{beenakker} for a model with
charging energy alone, and extended to the Universal Hamiltonian case
by Alhassid, Rupp, Kaminski, and Glazman\cite{alhassid1}. We follow
their notation here. The formula for conductance is\footnote{We differ
with them on an overall sign for this formula.}

 \beq
  G= {e^2\over \hbar k T}\sum_{ij} \tilde{P}_{i}^{(N)}f_{ij}\left(
  \Psi_{i}^{(N+1)}-\Psi_{i}^{(N)}+\eta_L\right) \Gamma^{L}_{ij}\label{G}
  \eeq
  where (i) $i$ and $j$ label many-body states with $N$ and $N+1$
  particles with energies $\epsilon^{(N+1) }_{j}$ and $ \epsilon^{(N) }_{i}$,
   (ii) $\epsilon_{ij}= \epsilon^{(N+1) }_{j} - \epsilon^{(N)
   }_{j}- \e_F -eV_g$ (where $\epsilon_F$ is the Fermi energy, chosen
   equal to zero and $V_g$ is the effective gate voltage), (iii)
   $\tilde{P}_{i}^{(N)}$
   is the equilibrium probability of being in state $i$ , (iv) $f_{ij}=f(\epsilon_{ij})$
   is the Fermi function, with $f(0)={1 \over 2}$, (v) $\eta_L$ is the fraction of the  bias
   voltage dropped at the left contact with $\eta_L+\eta_R=1$, (vi)
$\Gamma_{ij}^{L}$ is the width for many-body transition $i\to j$ at
the left contact
   when an electron jumps into the dot and (vii) $\Psi$ are functions
    defined by non-equilibrium probabilities in $P_{i}^{(N)}$ and $P_{j}^{(N+1)}$ to be in states $i$ and $j$ respectively in
    the presence of a
   bias voltage $V$
   \begin{eqnarray}
   P_{i}^{(N)}&=&\tilde{P}_{i}^{(N)}\left[ 1 + e V \beta
   \Psi_{i}^{(N)}\right]\label{38}\\
P_{j}^{(N+1)}&=&\tilde{P}_{j}^{(N+1)}\left[ 1 + e V \beta
   \Psi_{j}^{(N+1)}\right].\label{39}
   \end{eqnarray}
   The values of $\Psi$'s are determined by demanding that the
   $P$'s constitute a
   time-independent  solution to the rate equations and their normalization
   condition. The resulting equations are
   \begin{eqnarray}
   \sum_{j} f_{ij}\left[ (\Gamma_{ij}^{L}+\Gamma_{ij}^{R})
   (\Psi_{j}^{(N+1)}-\Psi_{i}^{(N)})\right. & &  \nonumber \\
   +\left. (\eta_{L}\Gamma_{ij}^{L}-
   \eta_{R}\Gamma_{ij}^{R})\right] &=&0\ \ \ \forall i \label{r1} \\
& & \nonumber  \\ \sum_{i} (1-f_{ij})\left[
(\Gamma_{ij}^{L}+\Gamma_{ij}^{R})
   (\Psi_{j}^{(N+1)}-\Psi_{i}^{(N)})\right. & & \nonumber \\
   +\left. (\eta_{L}\Gamma_{ij}^{L}-
   \eta_{R}\Gamma_{ij}^{R})\right] &=&0\ \ \ \forall j \label{r2}\\
   & & \nonumber \\
   \sum_i \tilde{P}_{i}^{(N)}\Psi_{i}^{(N)}+ \sum_j
   \tilde{P}_{j}^{(N+1)}\Psi_{i}^{(N+1)}&=&0\label{norm}.
   \end{eqnarray}

   Several comments are needed here\cite{beenakker,alhassid1}. While
   there seem to be $i+j+1$ equations for the $\Psi$'s, one of the
   first $i+j$
   equations is redundant, which is where the last equation, from normalization, comes to the rescue.
   The result, Eqn.(\ref{G}),  which seems to depend on $\eta$, in fact  does not.  Likewise the answer which seems to depend on the left
   lead in an asymmetric manner,  is not. (There is an equally
   asymmetric expression in terms of the right contact which gives
   the same result\cite{beenakker,alhassid1}.)  All this is exemplified by the  case where Eqns.
   (\ref{r1}, \ref{r2}) hold for each term in the sum, i.e.,
   \beq
    \Psi_{i}^{(N)}- \Psi_{j}^{(N+1)}= {\eta_L \Gamma_{ij}^{L}- \eta_R
    \Gamma_{ij}^{R}\over \Gamma_{ij}^{L}+  \Gamma_{ij}^{L}}\ \ \ \
    \forall i,j \label{detailed}
    \eeq
    in which case the expression Eqn. (\ref{G}), for $G$ reduces to
   \beq
  G= {e^2\over \hbar k T}\sum_{ij}
  \tilde{P}_{i}^{(N)}f_{ij}{\Gamma^{L}_{ij}\Gamma^{R}_{ij}\over
  \Gamma^{L}_{ij}+\Gamma^{R}_{ij}}.
  \label{G2}
  \eeq

Note that in the very-low-temperature case we considered with just
a total of three states, Eqns. (\ref{detailed} ) and (\ref{G2})
are applicable. \footnote{ In all cases we considered, either $j$
had just one value and  $i$ had two or {\em vice versa}. Suppose
$j$ had just one value. For the one redundant equation we are
allowed to  ignore, we choose the middle equation (that sums over
$i$). The first equation then gives Eqn. (\ref{detailed}) for each
$i$.} Our heuristic conductance calculations (and the definition
of $\barg$ in terms of $G$) followed from above and the fact that
$f(0)={1\over 2}$. However, in our numerical calculations at
temperatures comparable to $\alpha$ they do not hold and we solved
Eqs.(\ref{r1},\ref{r2},\ref{norm}) numerically.

\begin{figure}[]
\narrowtext \epsfxsize=3in\epsfysize=2.5in \hskip
0.3in\epsfbox{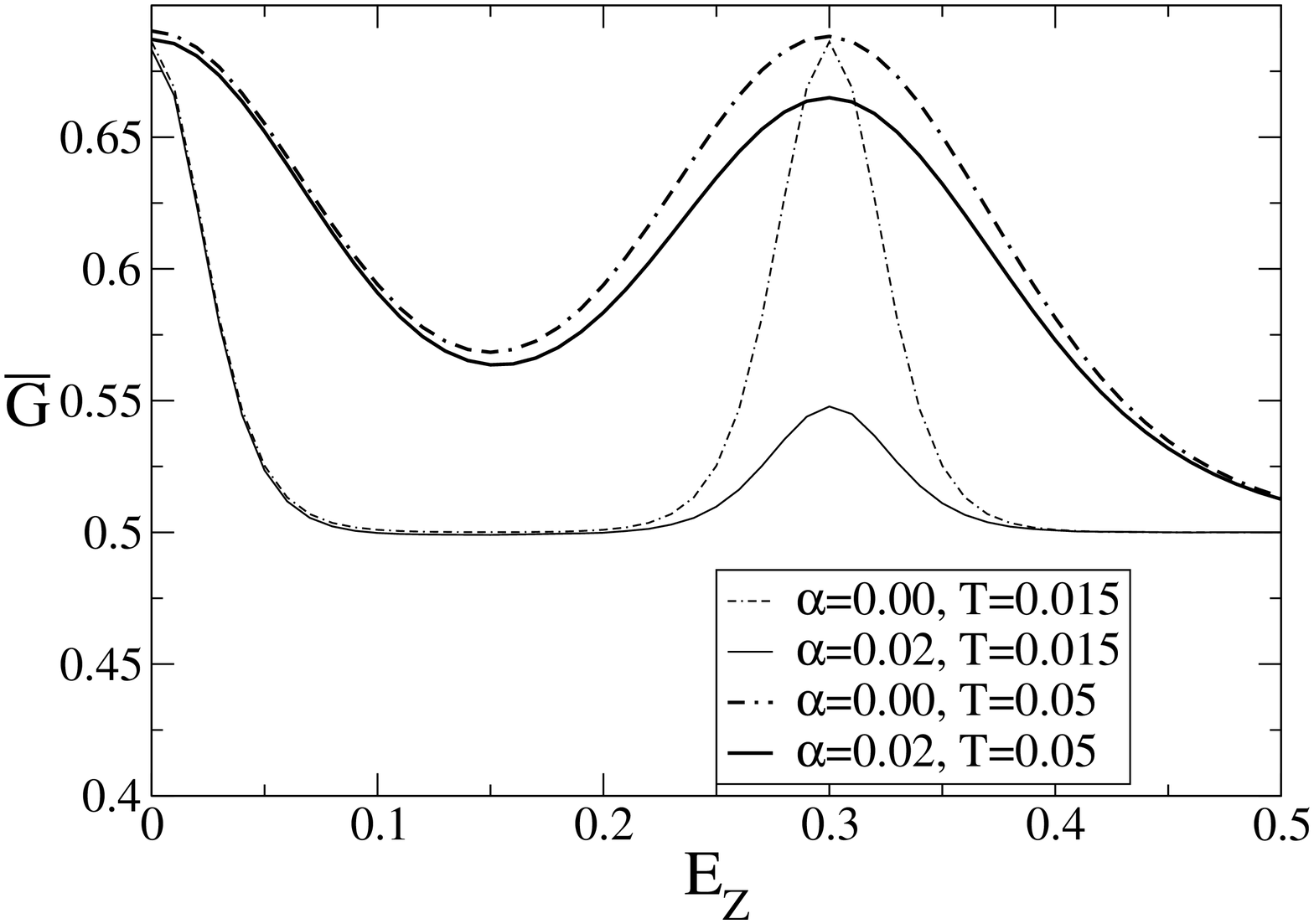} \vskip 0.15in \caption{Plot of $\barg$
versus $E_Z$ for the case $\barg^1=\barg^2$ and $l=1.1,r=.9$. Note
that when $\a=0$ (dotted lines), there is a bump at $E_{Z}^{*}$ as
$T\to 0$ which broadens with $T$, while with $\a \ne 0$, (solid
lines, non-zero spin-orbit coupling) there is hardly a bump as
$T\to 0$, but the bump grows with increasing $T$.} \label{g1}
\end{figure}

  We now display the result of solving the above equations for
  elevated temperatures using six states: two at $S=\half$ and
  four in the $S=0,1$ sector.
Since the graphs are functions of four variables,
$\barg^1,\barg^2,r,l$ we need to decide how to sample them here.
We have found that there are four main classes and give one from
each.\footnote{In choosing parameters for these graphs one cannot
 choose all the $\barg^1,\barg^2, r$ and $l$  arbitrarily. This is because even though every choice of
 the four $|\phi |^2$ will lead to positive values for $\barg^1,\barg^2, r$ and $l$, the converse is not true.
  Not  every choice of $\barg^1,\barg^2, r$ and $l$ has an inverse
 image with all   $|\phi |^2$' positive .
  To keep the left hand sides of  Eqns. (\ref{31},\ref{32})
  positive we must either satisfy    $l>r$,
 and $r <{\barg^1\over \barg^2}<l$, as in our choices above or a similar set of restriction for $r<l$. This is
not a concern in extracting $|\phi |^2$ from the experimental
$\barg^1,\barg^2, r$ and $l$,  they will automatically generate
the  sensible
 positive underlying values of
 $|\phi |^2$.}

\begin{figure}[]
\narrowtext \epsfxsize=3in\epsfysize=2.5in \hskip
0.3in\epsfbox{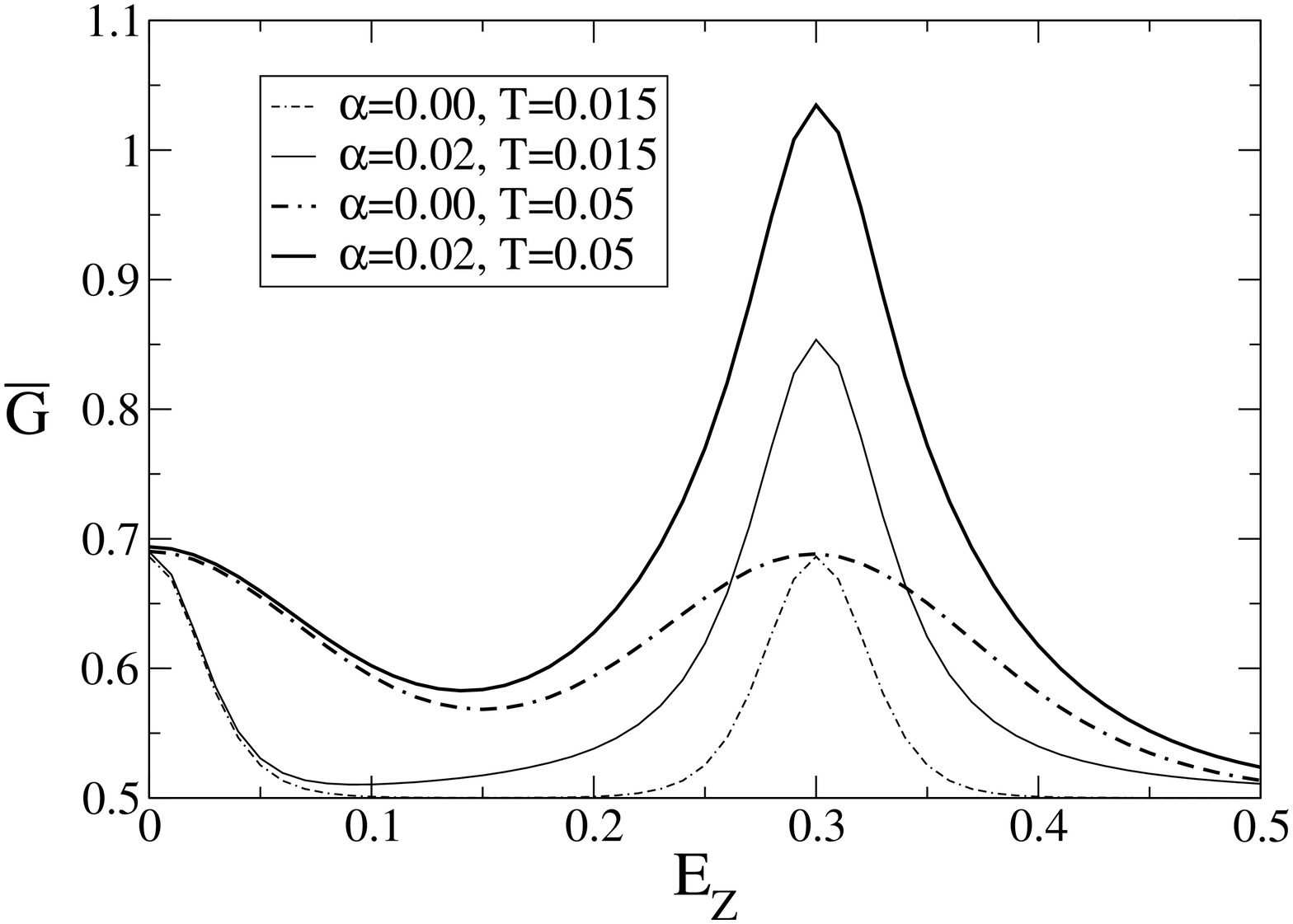} \vskip 0.15in \caption{Plot of $\barg$
versus $E_Z$ for the case $\barg^1=\barg^2$ and $l=4,r={1 \over
4}$. Note that when $\a=0$ (dotted lines) , there is a bump at
$E_{Z}^{*}$ as $T\to 0$ which broadens with $T$, while with $\a
\ne 0$, (solid lines) there is
 a prominent  bump as $T\to 0$, which  grows with increasing
$T$.} \label{g2}
\end{figure}

In Figures (\ref{g1},\ \ref{g2}) we consider the case
$\barg^1=\barg^2$. Recall that in this case, in the absence of
spin-orbit coupling ($\alpha=0$) there is a bump at $E_{Z}^{*}$ in
the reduced conductance $\barg^*$. This bump will broaden with
increasing $T$. When we turn on $\a$ there are two cases. If the
ratios of Eqs. (\ref{rdefn},\ref{ldefn}) satisfy $r\simeq l$ there
will be no bump in $\barg^{*}(\a)$ for $T<<\a$ and as we cross
$T>\a$, the $\a\ne0$ will look more and more like $\a=0$, as shown
in Figure (\ref{g1}). If however $r<<1$ and $l>>1$, the case
$\a\ne 0$ will have a bump at $E_{Z}^{*}$ even for $T<<\a$ that
will grow in height as $T$ increases as shown in Figure
(\ref{g2}). This bump will be robust even as the one at $E_Z=0$ is
smoothed out by temperature.

 Figures (\ref{g3}) and (\ref{g4}) consider the case of considerably  unequal
 $\barg$'s. We choose
for definiteness $\barg^1=2\barg^2$. The first case with $l=2.5$
and $r=1.5$ does not have a bump in $\barg^*$,
 or $\barg^*(\a)$ at low and high $T$, as shown in Figure
(\ref{g3}). However if $l$ are $r$ are different enough from each other (we
have chosen $ l=4.0, r=0.25$) there is a robust bump in $\barg^*(\a)$
that grows with increasing $T$
 as shown in Figure (\ref{g4}).

In summary, the most dramatic manifestations of $\a$ appear when
the $r$ and $l$ are quite different in size. It is also
interesting that the effect of $\a$ is felt at values of $T$
considerably higher than $\a$. This is because, as explained in
the paragraph below Eq. (\ref{symmcase}), {\it both} the states at
the avoided level crossing can and do  take advantage of large
level widths.

\begin{figure}[t]
\narrowtext \epsfxsize=3in\epsfysize=2.5in \hskip
0.3in\epsfbox{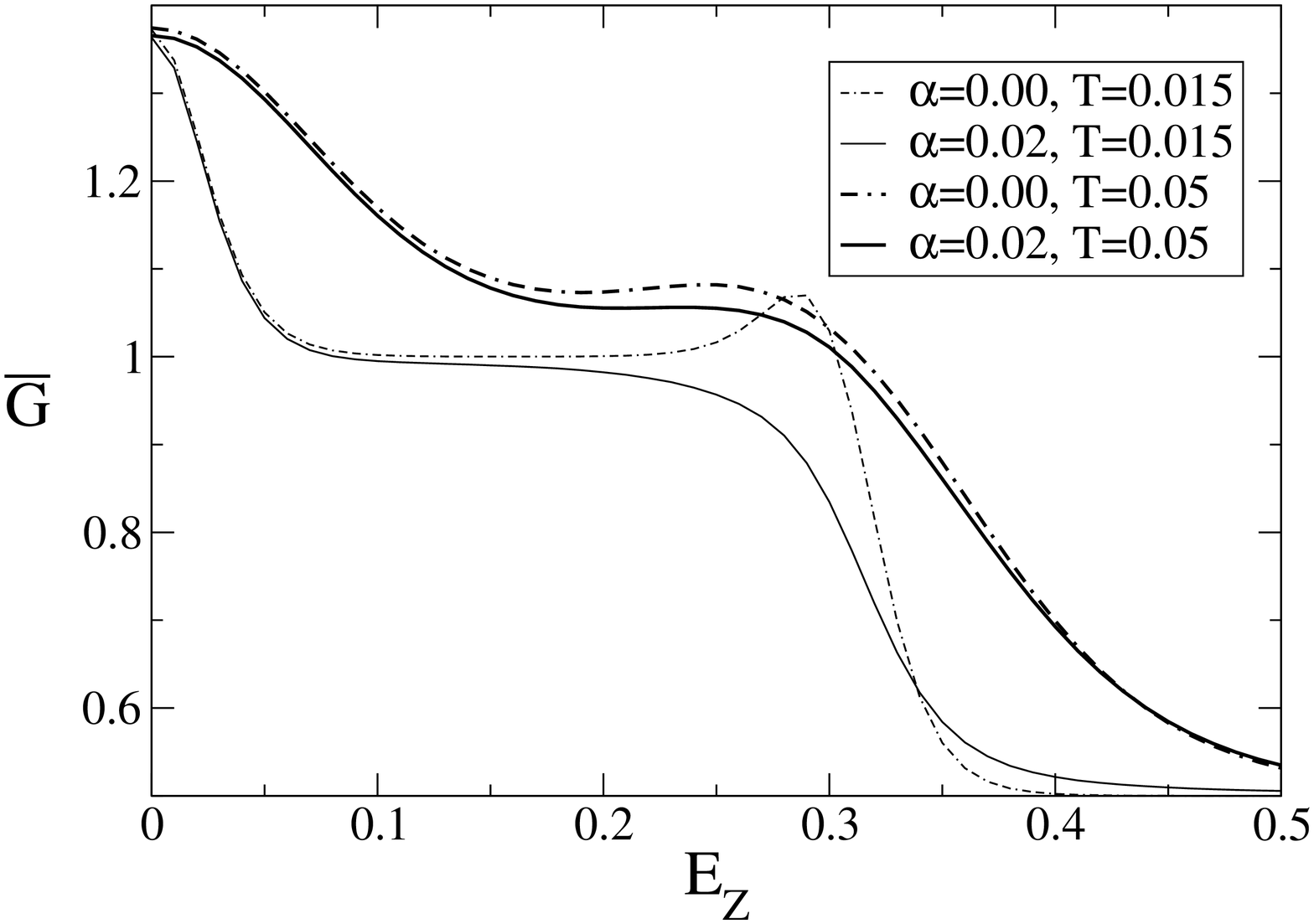} \vskip 0.15in \caption{Plot of $\barg$
versus $E_Z$ for the case $\barg^1=2\barg^2$ and $r=1.5 , l=2.5$.
There is no prominent bump with or without $\a$.} \label{g3}
\end{figure}

\begin{figure}[t]
\narrowtext \epsfxsize=3in\epsfysize=2.5in \hskip
0.3in\epsfbox{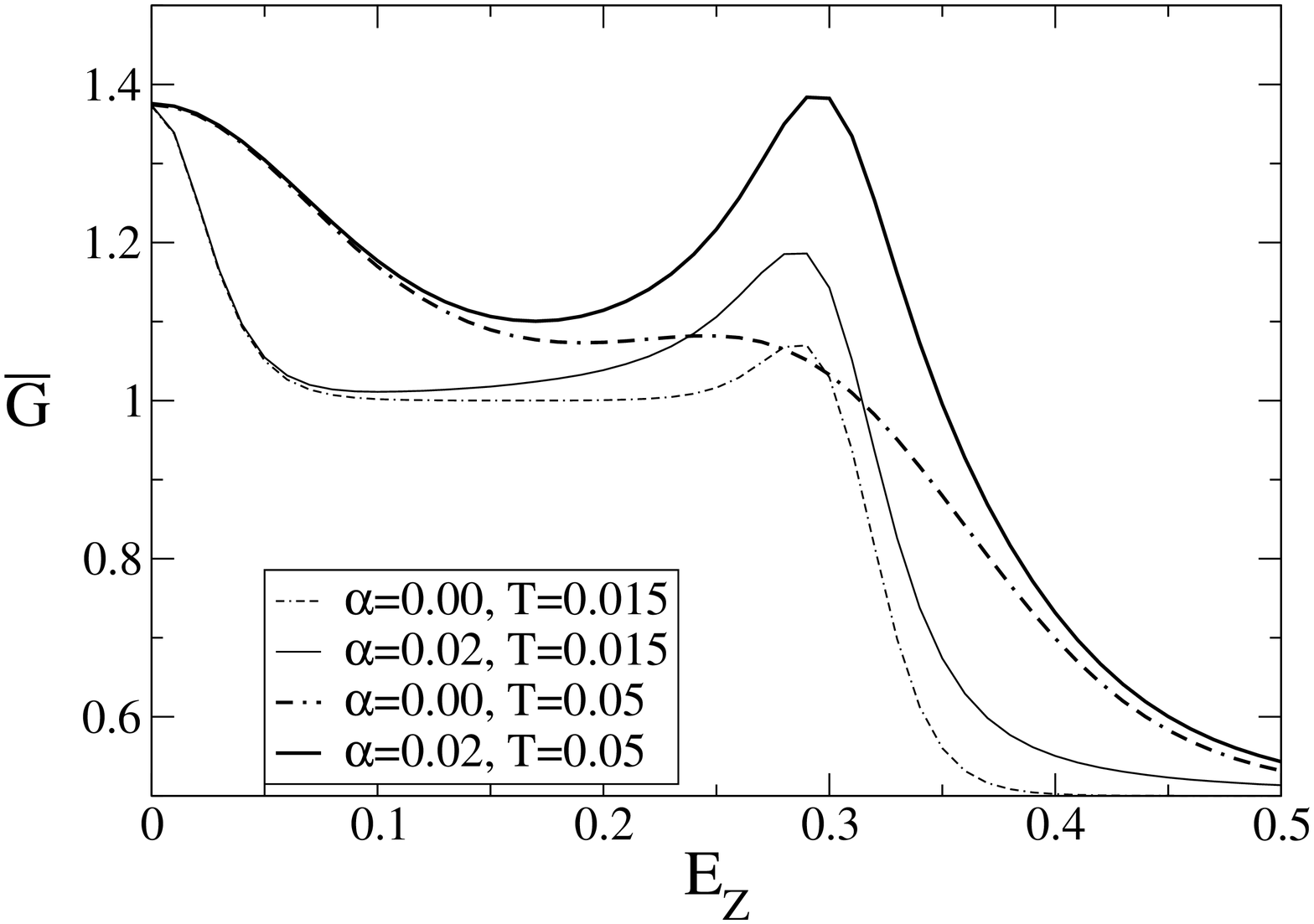} \vskip 0.15in \caption{Plot of $\barg$
versus $E_Z$ for the case $\barg^1=2\barg^2$ and $l=4,r={1 \over
4}$. There is no bump at $\a =0$ (dotted lines), but there is one
with $\a \ne 0$ (solid lines) which grows with $T$. } \label{g4}
\end{figure}

Finally we consider the effect of adding the term $\beta$ in the
spin-orbit interaction. A typical graph is found in Figure
(\ref{g5}) where even a $\beta$ twice as big as $\alpha$ is seen
to make little difference.

\begin{figure}[t]
\narrowtext \epsfxsize=3in\epsfysize=2.5in \hskip
0.3in\epsfbox{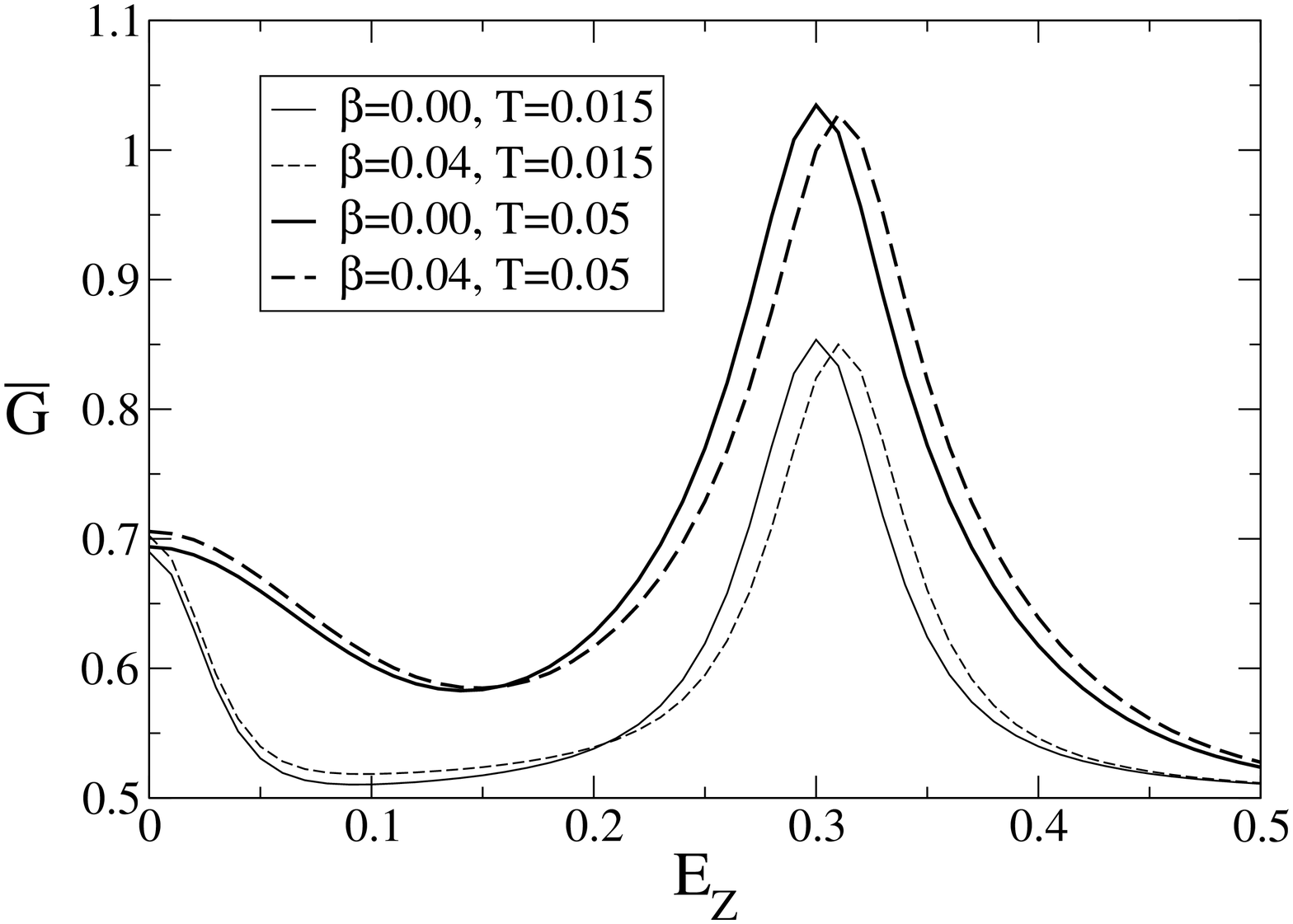} \vskip 0.15in \caption{Plot of $\barg$
versus $E_Z$ for the case $\barg^1=\barg^2$ and $l=4,r={1 \over
4}$. Both are at  at $\a =0.02$  but one has $\beta=0$ and the
other $\beta = 2 \a =.04$. Clearly $\beta$ makes little
difference.} \label{g5}
\end{figure}

  \subsection{Experimental checklist}

  Our paper began by looking for spin-orbit effects by starting
  with some input quantities and calculating experimentally
  measured numbers in terms of them. However, these input numbers,
  unlike say the electron mass or charge,  involved
  wave functions at the leads, energy gaps and so on,  and are not known {\em a priori} and the response of the
   system to turning on $\a$ is very sensitive to these quantities.

We suggest a possible road map below to deal with this, being fully
aware that our experimental colleagues will ultimately devise a better
solution.

\begin{itemize}
  \item Go to the lowest temperature possible.  \item By studying the
  $E_Z$ dependence of $V_g$\cite{potok} make sure the conductance peak
  describes the degeneracy of $(0,0)$ and $(\half, \half)$ which
  changes over to the degeneracy of $(1,1)$ and $(\half,
  \half)$. Since $\a\ne 0$, the rounded graph of Figure (\ref{vg})
  will be seen. Since $T<<\a$, the rounding is not due to thermal
  effects.\footnote{A caveat here is that the parallel magnetic field
  also has the effect of enhancing the confinement of the 2DEG in the
  $z$ direction\cite{jungwirth,potok}, leading to a quadratic
  dependence in all gate voltages. This will have to subtracted before
  our suggestion can be implemented.} \item By retracing the
  asymptotically
  linear segments in $V_g$ locate $E_{Z}^{*}$.  \item Since
  $E_{Z}^{*}=\delta - 2 J$, find $\delta$ if $J$ is known or use its
  average value of $.3 \bar{\delta}$.  \item Estimate the size of $\a$
  by comparing the difference of $V_g$'s between the two graphs at the
  cusp in Figure (\ref{vg}).
\item Measure $\barg (E_Z)$ and see if it looks like Figure
  (\ref{smoothg}). If so measure $\barg^1$ and $\barg^2$.
  \item By suppressing the overall tunnelling rate at the $R$ (right) and
  $L$(left)
  ends, (with a controllable barrier) measure $r$ and $l$, corresponding to the ratios of
  conductances. Bring tunnelling rates back to old value in the dot.
\item At $E_{Z}^{*}$ compare the predictions for $G^*(\a)$ the
value with $a\ne 0$ to the
  value without spin-orbit coupling, namely $\barg^*={2\over
  3}(\barg^1+\barg^2)$.
\item Raise $T$ and compare to theoretical predictions from the
  rate equations. All the parameters needed -- $|\phi_{0/1}(R/L)|^2, \a, \ \mbox{and} \ \delta$ -- are
  known by this point.
  \item Try to raise the value of $\a$ by
  turning on a inhomogeneities in the spin-orbit interaction using
  additional gates over the dot, following Brouwer {\it et
  al}\cite{BCH}.
  \end{itemize}

\section{Discussion and summary}

Spin-orbit couplings in quantum dots have been studied extensively in
the noninteracting limit in recent
years\cite{bert-ady,af,cremers,BCH,koneman,falko,ahmadian}. In systems
with homogeneous spin-orbit couplings, there is a new
symmetry\cite{af} which emerges at the Aleiner-Fal'ko (AF) scale
$\e^{AF}\simeq\lol^4 g\dbar$, corresponding to a $s_z$ symmetry in a
unitarily transformed basis\cite{af}. This symmetry is broken at a
scale $\lol^6 g\dbar$. For small quantum dots with $\lol\simeq 0.1$
and $g\simeq10$, one might therefore expect spin-orbit effects to be
utterly negligible.

Here we considered the  problem when a robust exchange term
$-J\bS^2$ of the Universal Hamiltonian is present. There are order
unity effects on the conductance at and near special degeneracy
points, where two different many-body ground states with equal
particle number but different total spin have the same energy.
{\em These effects have unique signatures that can be produced
only by the spin-orbit interaction.} We have investigated the
circumstances in which spin-orbit effects have the best prospect
of making their presence felt, and proposed ways to measure the
effect and compare it to theory. The ideal venue for studying the
phenomenon discussed here is in dots where $J$  has already
brought the $S=1$ state close to the $S=0$ state and one tunes
through the degeneracy by a  Zeeman coupling\cite{potok}. Our
results are robust under the addition of non-universal Landau type
interactions.

Our calculations, done in the orthogonal basis,  showed that in
the presence of $J$  spin-orbit terms enter the hamiltonian to
order $\lol$. We also saw how this phenomenon would appear in the
AF basis, in which the spin-orbit interaction has
been subsumed in the very choice of basis and a conserved quantity
$\sigma_{z}^{AF}$ emerges. In this basis the culprit is the
$-J\bS^2$ term which does not commute with $H_{AF}$, the
non-interacting AF hamiltonian. Upon unitary transformation, one
finds $-J\bS^2=-J(\bS^{AF}+\delta\bS^{AF})^2$, where
$\delta\bS^{AF}$ (which does not commute with $\sigma_{z}^{AF}$)
is order $\lol/\sqrt{g}$ for homogeneous spin-orbit coupling. We
have verified, by doing a calculation in the AF basis, that with
the nonzero $J$, the AF degeneracy is indeed lifted and the
spectrum coincides with what was found in the orthogonal basis.

Although the spin-orbit term appears to a lower order in ${L\over
\lambda}$ in the presence of $J$ than in the AF basis without $J$,
its parametric dependence on the Thouless number $g$ is very
different for homogeneous spin-orbit couplings. Whereas in the AF
basis a typical matrix element of the spin-orbit coupling is
$\lol^2\sqrt{g}$, in the original basis its typical size is
$\lol/\sqrt{g}$. The reason is evident from Eq.
(\ref{p=rdot})\cite{bert-ady}. In the original basis, the typical matrix elements
of spin-orbit coupling {\it averaged over the entire Thouless
band} are order $\lol\sqrt{g}$, but the matrix elements between levels
separated by $\dbar$ are of size $\lol/\sqrt{g}$.  In the AF
basis, the spin-orbit matrix elements are those of $(\br \times
\bp_{AF})$ which does not suffer from this suppression. This is why
Brouwer {\it et al}\cite{BCH} suggested that one could boost the
spin-orbit term by making its coefficient $1/\lambda$ space
dependent. In this case the matrix elements of the spin-orbit
coupling, even between neighboring states, would go parametrically
as $\lol\sqrt{g}$.

Throughout this paper we have restricted ourselves to low
temperatures $T\ll\dbar$, since that allows us to deal with only a
few many-body states. Let us briefly consider the effect of
raising $T$ to be of order $\dbar$ or higher, and considering the
effect on the conductance peak height
distribution\cite{alhassid3}. Excited states are populated at
higher $T$, and these will suffer stronger spin-orbit effects,
since the typical matrix elements of the spin-orbit coupling
increase with energy (Eq. (\ref{p=rdot2})). In parallel with this
increase, states of higher spin suffer larger spin-orbit effects,
even holding the matrix element constant. It will be interesting
to ask how this affects  the peak height distribution at moderate
temperatures, and in particular if it brings the theory into
better accord with experiments\cite{patel}.

We have concentrated here on small quantum dots in which
$\e^{AF}\ll\dbar$, but where $\lol\dbar/\sqrt{g}$ is still
accessible at low temperatures. Our philosophy has been to keep
$g$ constant, and perturb in $\lol$.

 For conceptual completeness,
let us consider the other limit, when
$\e^{AF}=\lol^4g\dbar\gg\dbar$. This regime would be accessible
for very large dots.  For noninteracting systems Aleiner, Fal'ko
and co-workers\cite{af,cremers} have made predictions for
conductance averages and fluctuations in open dots, which have
been confirmed experimentally\cite{zumbuhl}.

For closed dots with interacting electrons, one expects that at
energies very low compared to the AF scale $\e^{AF}$, the Universal
Hamiltonian has only Ising exchange interactions\cite{alhassid2},
since only $S_z^{AF}$ is conserved (see appendix).  As one increases
the energy, one expects to see {\it universal} crossover effects in,
for example, the spectral density of transverse spin
excitations\cite{murthy}. At energies much larger than $\e^{AF}$
orthogonal symmetry is restored. Both these calculations started with
the interaction term $-J(\bS^{AF})^2$, which, as we know, differs from
the correct starting point (see appendix), $-J\bS^2$. However, the
difference, which is at least first order in $\delta\bS^{AF}$,
vanishes as $g\to\infty$. In this limit, the previous
analyses\cite{alhassid2,murthy} are correct.

Finally, let us briefly touch upon the Stoner transition in the
presence of a small spin-orbit coupling. In the orthogonal class, one
sees a series of metamagnetic transitions\cite{univ-ham,hu-reviews}
with states of successively higher conserved spin as $J\to \dbar$,
with the bulk Stoner transition being their accumulation point. In the
presence of spin-orbit interactions, for any finite $g$ these steps in
$S$ are rounded, and the transition will become a regular second-order
quantum phase transition, where the order parameter is not conserved
(as we have seen above, even $S_z^{AF}$ is not conserved for finite
$g$). On the other hand, if one works in the universal limit
$g\to\infty$ while keeping $\e^{AF}/\dbar$ fixed, there will be a
sequence of metamagnetic transitions between states with increasing
conserved $S_z^{AF}$.  For very large but finite $g$ one can expect
the steps of $S_z^{AF}$ to be rounded and merge into a smooth curve as
$J\to\dbar$. We hope to present a full analysis of the different
regimes in a future publication.

To conclude, we have found that in the presence of a robust
exchange interaction $J$, one can, by  tuning  the Zeeman
coupling\cite{potok}, readily expose order-unity effects of the
spin-orbit coupling on the conductance at and near a many-body
degeneracy. The effects have signatures that allow one to trace
them  back to the spin-orbit coupling. To access these effects
experimentally one needs to  reduce the temperature to be of the
order of the spin-orbit coupling strength. Our results are
unaffected by a modest amount of spin-conserving non-universal
interactions. We have offered a proposal for a sequence of
experiments to determine all the relevant parameters, that is, the
nearest neighbor level spacing and the single-particle wave
functions at both leads, which should enable experimentalists to
test the predictions of this paper. In typical small dots with
$\dbar\simeq 0.1meV$, $\lol\simeq0.1$ and $g\simeq10$, we hope
that measurements to test the theory will be achievable in the
near future.

\section{Acknowledgements}

We are grateful to Leonid Glazman, David Goldhaber-Gordon, Bert
Halperin, and Karyn LeHur for discussions. Special thanks go to Yoram
Alhassid for countless illuminating conversations and several thorough
readings of the manuscript, and to the Aspen Center for Physics, where
this paper was conceived.  RS acknowledges the National Science
Foundation for DMR 0354517.

\section{Appendix A}

Here we show that adding the $-J\bS^2$ is the correct effective theory
at the Thouless scale $E_T$ as long as $\e_{AF}\ll E_T$, and discuss the
possible low-energy theories for energies much less than $\e_{AF}$.

Our reasoning is based on renormalization group (RG), which is an
unbiased method for revealing the relevant and irrelevant terms, and
is as follows: As is well-known, fermionic RG\cite{rg-shankar} in the
clean system leads to Landau Fermi liquid theory\cite{agd} on a scale
$E_{FLT}$ much smaller than the Fermi energy. Since $E_T$ can be made
as small as we want by increasing the size $L$, we can always achieve
the condition $E_T\ll E_{FLT}$. Therefore the correct starting point
at the Thouless scale is a Fermi liquid theory. This theory is
parameterized by Landau interaction parameters\cite{agd} $u_m$
$m=0,1,\cdots$ in the spin singlet and triplet channels. For $m=0$ the
Landau parameters are (up to a factor of $\dbar$) none other than the
charging and spin exchange energies. Now consider carrying out RG in
the original disordered basis with both the two-body Fermi liquid
interaction and in addition a spin-orbit coupling. Without the
spin-orbit coupling, this has been done by one of the authors and
Harsh Mathur\cite{mm}. The results are: (i) Any two-body interaction
which is a conserved quantity squared will not flow. This includes the
charging interaction and the $-J\bS^2$ Stoner interaction in the
Universal Hamiltonian. (ii) Any other two-body interaction will flow,
and flow to zero at weak coupling. This means that one can neglect the
non-$s$-wave Landau interactions since they are irrelevant, and allows
us to start with the universal hamiltonian $-J\bS^2$ coupling at the
Thouless energy. This argument also shows that higher channel
spin-rotation-invariant Landau interactions do not affect the results
of this paper, as long as they are weak.

Coming now to the added spin-orbit coupling, this will grow in RG as the
cutoff scale decreases, indicating that it is relevant for low-energy
physics. At some cutoff scale the size of the spin-orbit coupling will rival
that of the cutoff. This scale should be identified as
$\e_{AF}$. Below this scale, one cannot treat the spin-orbit coupling
perturbatively, but should start with exact states which incorporate
it. In the present paper, we are working in the case
$\e_{AF}\ll\dbar$, allowing us to treat the spin-orbit coupling
perturbatively.

For the case $\e_{AF}\gg\dbar$, neglecting the terms of order $\lol^2
\e_{AF}$ which break this $S_{z,AF}$ symmetry, there is a separate
universal hamiltonian regime at energies much smaller than
$\e_{AF}$. This regime was identified by Alhassid and
Rupp\cite{alhassid2} and has only that part of the interaction which
commutes with the kinetic term $-J_z S_{z,AF}^2$. One can obtain the
same result by RG, this time diagonalizing the kinetic term with
spin-orbit, and realizing that since the in-plane parts of the
original Stoner interaction are not conserved, they do flow, and flow
moreover to zero leaving behind the $-J_z S_{z,AF}^2$.

\end{document}